\documentclass[a4paper,12pt]{article}

\usepackage{amsmath}
\usepackage[psamsfonts]{amssymb}
\usepackage{rsfs}
\usepackage{bm}

\usepackage{cite}
\usepackage[dvips]{graphicx}
\usepackage{color}

\begin{document} 

\begin{titlepage}

\baselineskip 10pt
\hrule 
\vskip 5pt
\leftline{}
\leftline{Chiba Univ./KEK Preprint
          \hfill   \small \hbox{\bf CHIBA-EP-157}}
\leftline{\hfill   \small \hbox{\bf KEK Preprint 2006-1}}
\leftline{\hfill   \small \hbox{hep-th/0604006}}
\leftline{\hfill   \small \hbox{August 2006}}
\vskip 5pt
\baselineskip 14pt
\hrule 
\vskip 1.0cm
\centerline{\Large\bf 
Glueball mass  from 
} 
\vskip 0.3cm
\centerline{\Large\bf  
quantized knot solitons  
}
\vskip 0.3cm
\centerline{\Large\bf  
and gauge-invariant gluon mass
}

\vskip 0.7cm

\centerline{{\bf 
Kei-Ichi Kondo,$^{\dagger,\ddagger,{1}}$  Akihito Ono,$^{\ddagger,{2}}$ Akihiro Shibata,$^{\flat,,{3}}$ 
}}  
\vskip 0.3cm
\centerline{{\bf 
Toru Shinohara,$^{\ddagger,{4}}$and Takeharu Murakami$^{\ddagger,{5}}$
}}  
\vskip 0.5cm
\centerline{\it
${}^{\dagger}$Department of Physics, Faculty of Science, 
Chiba University, Chiba 263-8522, Japan
}
\vskip 0.3cm
\centerline{\it
${}^{\ddagger}$Graduate School of Science and Technology, 
Chiba University, Chiba 263-8522, Japan
}
\vskip 0.3cm
\centerline{\it
${}^{\flat}$Computing Research Center, High Energy Accelerator Research Organization (KEK),  
}
\vskip 0.3cm
\centerline{\it
Tsukuba 
305-0801, 
\& 
Graduate Univ. for Advanced Studies (Sokendai),
 Japan
}
\centerline{\it
}
\vskip 0.5cm

\begin{abstract} 
We propose an approach which enables one to obtain simultaneously the glueball mass and the gluon mass in the gauge-invariant way to shed new light on the mass gap problem in Yang-Mills theory.
First, we point out that the Faddeev (Skyrme--Faddeev-Niemi) model  can be induced through the gauge-invariant vacuum condensate of mass dimension two from SU(2) Yang-Mills theory. 
Second, we obtain the glueball mass spectrum by performing the collective coordinate quantization of the topological knot soliton in the Faddeev model.  
Third, we demonstrate that a relationship  between the glueball mass and the gluon mass is obtained, since the gauge-invariant gluon mass is also induced from the relevant vacuum condensate.  
Finally, we determine physical values of two parameters in the Faddeev model and give an estimate of the relevant vacuum condensation in Yang-Mills theory. 
Our results indicate that the Faddeev model can play the role of a low-energy effective theory of the quantum SU(2) Yang-Mills theory.

\end{abstract}

Key words:   glueball mass, vacuum condensate,  Faddeev model, knot soliton, magnetic monopole, quark confinement,

PACS: 12.38.Aw, 12.38.Lg 
\hrule  
\vskip 0.1cm
${}^1$ 
  E-mail:  {\tt kondok@faculty.chiba-u.jp}
  
${}^2$ 
  E-mail:  {\tt a.ono@graduate.chiba-u.jp}
  
${}^3$ 
  E-mail:  {\tt akihiro.shibata@kek.jp}
  
${}^4$ 
  E-mail:  {\tt sinohara@graduate.chiba-u.jp}
  
${}^5$ 
  E-mail:  {\tt tom@cuphd.nd.chiba-u.ac.jp}

\par 
\par\noindent


\vskip 0.5cm

\newpage
\pagenumbering{roman}




\end{titlepage}


\pagenumbering{arabic}

\baselineskip 14pt
\section{Introduction}

It is widely accepted that the quantized Yang-Mills theory, i.e., quantum gluodynamics (QGD) is the fundamental theory for the strong force.  In the high-energy region, the perturbative expansion in the coupling constant gives an efficient method for calculations by virtue of the ultraviolet asymptotic freedom. In the low-energy region, however, some non-perturbative methods are needed to tame the strong coupling problem.  It is rather difficult to perform the analytical calculation to derive non-perturbative results keeping the gauge invariance manifest. 
In such a case, it is sometimes useful to investigate the corresponding low-energy effective theory (LEET) instead of tackling the fundamental theory itself. 

In this paper, we demonstrate that the Faddeev (or Faddeev--Niemi) model \cite{Faddeev75} can be used as a realistic LEET of SU(2) Yang-Mills theory by deriving the  gauge-invariant  glueball mass and  gluon mass of the original Yang-Mills theory from the Faddeev model.  
In fact, it was suggested in a previous paper \cite{Kondo04} that the Faddeev model is obtained as a low-energy effective theory of QGD by performing the non-linear change of variables (NLCV) of the gluon field $\mathscr{A}_\mu$ \cite{KMS06} and assuming the existence of novel gauge-invariant vacuum condensation $\langle \mathbb{X}_\mu^2 \rangle_{\rm YM}$ of mass dimension two
\footnote{
This vacuum condensation corresponds to a gauge-invariant version  of the on-shell BRST-invariant vacuum condensation of mass dimension two in the modified Maximal Abelian gauge and the generalized Lorentz gauge proposed in \cite{Kondo01}.
The existence of $\langle \mathbb{X}_\mu^2 \rangle_{\rm YM}\ne 0$ in the original Yang-Mills theory has been examined in numerical simulations \cite{KKMSS05}.  The analytical treatment of this condensate will be given elsewhere and is not discussed  in this paper.
}
 \cite{KKMSS05}. 
The relationship between Yang-Mills theory and the Faddeev model is similar in spirit to that between QCD and the Skyrme model which expresses baryons as topological solitons of the meson field \cite{ANW83}. 

The relevant NLCV for SU(2) gluon field $\mathscr{A}_\mu(x)$ is represented by introducing the unit vector field $\vec{n}(x)=(n_A(x)),  (A=1,2,3)$ as  
\footnote{
This form of NLCV was known as the CFN decomposition \cite{Cho80,FN98,Shabanov99} proposed by Cho \cite{Cho80} and Faddeev--Niemi \cite{FN98}. In \cite{KMS06}, three of the authors have given a new viewpoint of the CFN decomposition on which this paper is based. 
}
\begin{align}
\mathscr{A}_\mu(x)
 =& c_\mu(x) \bm{n}(x)
  + i g^{-1}[ \bm{n}(x), \partial_\mu \bm{n}(x)  ]
  +\mathbb X_\mu(x) ,
\nonumber\\
 &  2{\rm tr}(\bm{n}(x)   \bm{n}(x) ) = 1   , 
\quad {\rm tr}(\bm{n}(x)  \mathbb{X}_\mu(x)) = 0 ,
\end{align}
where we have used the  $su(2)$  valued field:
$
 \bm{n}(x)  := n_A(x) T_A
$,
$
 \mathbb{X}_\mu(x)  := X_\mu^A(x) T_A
$,
$
 T_A = \frac12 \sigma_A ,
$
($\sigma_A $: Pauli matrices)
and 
$c_\mu$ and 
 $\mathbb{X}_\mu$ are specified by $\bm{n}$ and $\mathscr{A}_\mu$ as 
\begin{align}
c_\mu(x) &= 2{\rm tr}({\bm n}(x) \mathscr{A}_\mu(x)) , 
\quad 
\mathbb X_\mu(x) = -i g^{-1}[{\bm n}(x),  D_\mu[\mathscr{A}]{\bm n}(x)] .
\label{def:X}
\end{align}

The existence of gauge-invariant vacuum condensate of mass dimension two $\langle \mathbb{X}_\mu^2 \rangle_{\rm YM}$ has the following implications. 

(1) The gauge-invariant gluon mass is deduced from the four-gluon interaction $-\frac14 g^2 (\mathbb{X}_\mu \times \mathbb{X}_\nu)^2$.  For example, a simple  self-consistent treatment in the mean field or Hartree-Fock appoximation leads to the replacement: 
\begin{align}
 & -\frac{1}{4}(g \mathbb{X}_\mu \times \mathbb{X}_\nu) \cdot (g \mathbb{X}^\mu \times \mathbb{X}^\nu) 
 \nonumber\\
\to & \frac{1}{2}g^2 \mathbb{X}^A_\mu \left[\left\langle -\mathbb{X}^2_\rho \right\rangle_{\rm YM} \delta^{AB} - \left\langle -\mathbb{X}^A_\rho \mathbb{X}^B_\rho \right\rangle_{\rm YM} \right] \mathbb{X}^B_\mu 
=   \frac{1}{2} M_X^2 \mathbb{X}_\mu \cdot \mathbb{X}_\mu ,
 \quad
M_X^2 \propto g^2 \left\langle -\mathbb{X}^2_\rho \right\rangle_{\rm YM} .
\end{align} 
Therefore, $X$-gluons (off-diagonal gluons) acquire the mass 
$M_X  \propto \sqrt{g^2 \left\langle -\mathbb{X}^2_\rho \right\rangle_{\rm YM}}$
and eventually decouple in the low-energy region below this scale.  After integrating out the $X$-gluon field, the resulting LEET is expected to be written in terms of the vector field $\vec{n}(x)$ alone. This LEET will agree with the Faddeev model.%
\footnote{
There are other approaches for deriving the Faddeev model from Yang-Mills theory. See e.g., \cite{LN99,Gies01}. 
}

(2) The (gauge-invariant) glueball mass can be derived from the Faddeev model  with the Lagrangian density:
\begin{align}
  \mathscr{L}_{\rm F} 
  =  \frac{1}{2} \Lambda_{\rm F}^2 (\partial_\mu \vec{n})^2 - \frac{1}{4} C_4  [\vec{n} \cdot (\partial_\mu \vec{n} \times \partial_\nu \vec{n})]^2  ,
\end{align}
where the bare (tree) values of the parameters are given by \cite{Kondo04}
\begin{align}
 \Lambda_{\rm F} = \sqrt{\langle -\mathbb{X}_\mu^2 \rangle_{\rm YM}}, \quad C_4 = 1/g^2 . 
 \label{Faddeev-model-parameter}
\end{align}
Then the two parameters $\Lambda_{\rm F}$ and $C_4$ in the Faddeev model are gauge invariant in the original Yang-Mills theory under this identification and
 the physical quantities calculated in the Faddeev model are automatically  guaranteed to be gauge-invariant in the original Yang-Mills theory.  

(3) Quark confinement is expected to follow from the non-vanishing string tension $\sigma$ (as a coefficient of the linear quark-antiquark potential $V(r)=\sigma r$) which is proportional to $\langle \mathbb{X}_\mu^2 \rangle_{\rm YM}$:
\begin{align}
 \sigma = {\rm const.} \langle -\mathbb{X}_\mu^2 \rangle_{\rm YM} .
\end{align}
There exists an intimate relationship between the glue--knot and the magnetic monopole which is  a basic ingredient in the dual superconductor picture for quark confinement.  This issue will be discussed in a subsequent paper.

Thus, our approach presents an efficient framework for understanding the mass gap in Yang-Mills theory and quark confinement in QCD. 
In this paper,  as a step along this line, we derive glueball mass spectra by performing the collective coordinate quantization of knot soliton solution in the Faddeev model.  
Glueballs are identified with knot solitons and their excitations. 
The knot soliton is a topological soliton with non-vanishing Hopf index $Q_H$ whose existence is suggested  by the non-trivial Homotopy group $\pi_3(S^2)=\mathbb{Z}$. 
We can determine the two parameters $\Lambda_{\rm F}$ and $C_4$ of the Faddeev model by using two glueball masses $M_{0^{++}}$ and $M_{2^{++}}$ for $0^{++}$ and $2^{++}$  (obtained by numerical simulations on a lattice) as inputs to predict glueball masses other than $0^{++}$ and $2^{++}$. 
Finally, we estimate the gluon mass $M_X$ and estimate the vacuum condensation $\langle -\mathbb{X}_\mu^2 \rangle_{\rm YM}$ of the original Yang-Mills theory.

\section{Classical knot soliton solution}

First, we obtain the knot soliton as  a static and finite-energy solution $\vec{n}(\vec{x})$ of the Faddeev model with  
the energy  given by
\begin{align}
  E =& \int_{\mathbb{R}^3} d^3x  \left\{ \frac{1}{2} \Lambda_{\rm F}^2 (\partial_j \vec{n})^2 + \frac{1}{4} C_4  [\vec{n} \cdot (\partial_j \vec{n} \times \partial_k \vec{n})]^2  \right\} \quad (j,k=1,2,3) .
\end{align}
The energy $E_{\Lambda_{\rm F},C_4}[\vec{n}(\vec{x})]$ as a functional of $\vec{n}(\vec{x})$ for arbitrary values of the two parameters  $\Lambda_{\rm F}$ and $C_4$ obeys the scaling relation:
\begin{align}
 E_{\Lambda_{\rm F},C_4}[\vec{n}(\vec{x})]  =\Lambda_{\rm F} C_4^{1/2} E_{1,1}[\vec{n}(\Lambda_{\rm F}^{-1} C_4^{1/2}\vec{x})] .
 \label{E-scaling}
\end{align} 
Once the solution $n^{*}$ minimizing $E_{1,1}$ is known, therefore, the solution corresponding to $E_{\Lambda_{\rm F},C_4}$ is given by 
\begin{align}
 \vec{n}_{\Lambda_{\rm F},C_4}(\vec{x})  = \vec{n}_{*}(\Lambda_{\rm F} C_4^{-1/2} \vec{x}) ,
\end{align} 
and the mass of the knot soliton for arbitrary parameters obeys the relation 
\begin{align}
 M_{\Lambda_{\rm F},C_4} = E_{\Lambda_{\rm F},C_4}[\vec{n}_{\Lambda_{\rm F},C_4}] = \Lambda_{\rm F} C_4^{1/2} E_{1,1}[\vec{n}_{*}]
 =  \Lambda_{\rm F} C_4^{1/2} M_{*} .
\end{align} 
In what follows, $*$ denotes the quantity calculated at $\Lambda_{\rm F}=1$ and $C_4=1$.

We adopt a simplified version \cite{HS98}  of the toroidal ansatz \cite{GH97}:
\begin{align}
  \vec{n} :=  (n_1, n_2, n_3) =& \left(\frac{2f(\eta)}{f^2(\eta)+1} \cos (m\xi-n\varphi),\frac{2f(\eta)}{f^2(\eta)+1} \sin (m\xi-n\varphi),\frac{f^2(\eta)-1}{f^2(\eta)+1} \right) ,  
  \nonumber\\
 & (m, n=  \pm 1, \pm 2, \cdots) ,
 \label{toroidal}
\end{align}
where $m$ and $n$ are non-zero integers,  and  the toroidal coordinate
$(\eta, \xi, \varphi)$ is given by
\begin{align}
  x =& a \frac{\sinh \eta \cos \varphi}{\cosh \eta - \cos \xi} , \quad
  y = a \frac{\sinh \eta \sin \varphi}{\cosh \eta - \cos \xi} , \quad
  z = a \frac{\sin \xi}{\cosh \eta - \cos \xi}, 
  \nonumber\\
&  0 \le \eta < \infty, \quad -\pi \le \xi \le \pi, \quad  0 \le \varphi < 2\pi .
\end{align}

\begin{figure}[htbp]
\begin{center}
\includegraphics[height=3.5cm]{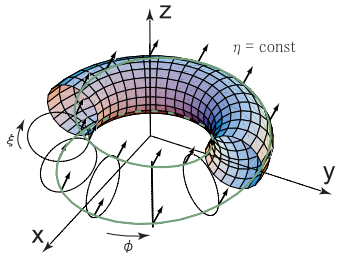}
\caption{\small 
The Hopfion configuration with 
$Q_H=2$ ($m=1, n=2$) is indicated by a set of arrows on the torus specified by a toroidal coordinates. 
}
\label{fig:Hopfion-torus2}
\end{center}
\end{figure}

It is easy to understand that the Hopf topological charge of the Hopfion
(static soliton solution with non-zero Hopf topological charges) under this ansatz is given by 
\begin{align}
  Q_H = mn   \quad (m, n=  \pm 1, \pm 2, \cdots), 
\end{align}
since the Hopf charge is equivalent to the linking number of two circles  (in $S^3$) obtained as preimages of two distinct points in the target space $S^2$. 
See Fig.~\ref{fig:Hopfion-torus2}. 


We consider the discrete and continuous symmetries of the color field $\vec{n}(\vec{x})$ under the toroidal Ansatz (\ref{toroidal}) characterized by indices $n$ and $m$ related to the Hopf charge $Q_H=nm$. 

\underline{Planar reflection} with respect to
\begin{enumerate}
\item[(a)] $yz$ plane: $x \rightarrow -x \Longleftrightarrow \varphi \rightarrow \pi - \varphi$
\begin{subequations}
\begin{align}
 n_1(-x,y,z)_{m,n} =& (-1)^{n} n_1(x,y,z)_{m,-n}=(-1)^{n} n_1(x,y,z)_{-m,n} ,
 \\
 n_2(-x,y,z)_{m,n} =& (-1)^{n} n_2(x,y,z)_{m,-n}=(-1)^{n+1} n_2(x,y,z)_{-m,n} ,
 \\
 n_3(-x,y,z)_{m,n} =&  n_3(x,y,z)_{m,n} ,
\end{align}
\end{subequations}
\item[(b)] $xz$ plane: $y \rightarrow -y \Longleftrightarrow \varphi \rightarrow - \varphi$
\begin{subequations}
\begin{align}
 n_1(x,-y,z)_{m,n} =&  n_1(x,y,z)_{m,-n}= n_1(x,y,z)_{-m,n} ,
 \\
 n_2(x,-y,z)_{m,n} =&   n_2(x,y,z)_{m,-n}= - n_2(x,y,z)_{-m,n} ,
 \\
 n_3(x,-y,z)_{m,n} =&  n_3(x,y,z)_{m,n} ,
\end{align}
\end{subequations}

\item[(c)] $xy$ plane: $z \rightarrow -z \Longleftrightarrow \xi \rightarrow - \xi $
\begin{subequations}
\begin{align}
 n_1(x,y,-z)_{m,n} =&  n_1(x,y,z)_{m,-n}= n_1(x,y,z)_{-m,n} ,
 \\
 n_2(x,y,-z)_{m,n} =& - n_2(x,y,z)_{m,-n}= n_2(x,y,z)_{-m,n} ,
 \\
 n_3(x,y,-z)_{m,n} =&  n_3(x,y,z)_{m,n} ,
\end{align}
\end{subequations}
\end{enumerate}
where $n_3(x,y,z)_{m,n}=n_3(x,y,z)_{-m,n}=n_3(x,y,z)_{m,-n}=n_3(x,y,z)_{-m,-n}$.

Composing these two at a time yields the symmetries associated with 

\underline{Spacial rotations by $\pi$ radians} about 
\begin{enumerate}
\item[(i)] $x$-axis (=(b)$ \times $(c))
\begin{align}
 \bm{n}(x,-y,-z)_{m,n} = 
 \begin{cases}
 - \sigma_2 \bm{n}(x,y,z)_{m,n} \sigma_2 
= - e^{i\pi \sigma_2/2} \bm{n}(x,y,z)_{m,n} e^{-i\pi \sigma_2/2}  &(n=\text{even})  \\
   - \sigma_2 \bm{n}(x,y,z)_{m,n} \sigma_2
= - e^{i\pi \sigma_2/2} \bm{n}(x,y,z)_{m,n} e^{-i\pi \sigma_2/2}  & (n=\text{odd}) 
 \end{cases} ,
 \label{x}
\end{align} 

\item[(ii)]   $y$-axis (=(a)$ \times $(c))
\begin{align}
 \bm{n}(-x,y,-z)_{m,n} =
 \begin{cases}
 - \sigma_2 \bm{n}(x,y,z)_{m,n} \sigma_2
= - e^{i\pi \sigma_2/2} \bm{n}(x,y,z)_{m,n} e^{-i\pi \sigma_2/2}  &(n=\text{even})  \\
   - \sigma_1 \bm{n}(x,y,z)_{m,n} \sigma_1
= - e^{i\pi \sigma_1/2} \bm{n}(x,y,z)_{m,n} e^{-i\pi \sigma_1/2}  & (n=\text{odd}) 
 \end{cases} ,
 \label{y}
\end{align} 

\item[(iii)]   $z$-axis (=(a)$ \times $(b))
\begin{align}
 \bm{n}(-x,-y,z)_{m,n} =
 \begin{cases}
 \bm{n}(x,y,z)_{m,n} &(n=\text{even})  \\
   \sigma_3 \bm{n}(x,y,z)_{m,n} \sigma_3 
= e^{i\pi \sigma_3/2} \bm{n}(x,y,z)_{m,n} e^{-i\pi \sigma_3/2}  & (n=\text{odd}) 
 \end{cases} ,
 \label{z}
\end{align} 
\end{enumerate}
where we have used 
$\sigma_j \sigma_k \sigma_j=-\sigma_k (j\ne k), \sigma_k (j=k)$ (no sum over $j$) following from $\sigma_j \sigma_k = \delta_{jk}\bm{1} + i \epsilon_{jk\ell} \sigma_\ell$,
and also the relation 
$e^{\pm i\pi \sigma_j/2}=\bm{1} \cos \frac{\pi}{2} \pm i \sigma_j \sin \frac{\pi}{2}= \pm i \sigma_j$.
Making the product of all three rotations (i),(ii) and (ii) recovers the original configuration, as a trivial cross check. 

Finally, we consider Parity transformation. 
Making the product of all three refections (a),(b) and (c) yields

\underline{Parity}:
\begin{align}
 & \bm{n}(-x,-y,-z)_{m,n} 
\nonumber\\
=&
 \begin{cases}
 \bm{n}(x,y,z)_{-m,n} 
= - e^{i\pi \sigma_2/2} \bm{n}(x,y,z)_{m,-n} e^{-i\pi \sigma_2/2}
&(n=\text{even}) \\
    e^{i\pi \sigma_3/2} \bm{n}(x,y,z)_{-m,n} e^{-i\pi \sigma_3/2} 
=- e^{i\pi \sigma_1/2} \bm{n}(x,y,z)_{m,-n} e^{-i\pi \sigma_1/2} 
 & (n=\text{odd}) 
 \end{cases} .
\end{align} 

Now we proceed to obtain the solution of the Faddeev model. 
After performing angular integrations over $\varphi$ and $\xi$ under the toroidal ansatz, we find that the energy functional reduces to 
\begin{align}
 E(\Lambda_{\rm F}, C_4;a) =&  \Lambda_{\rm F}^2 a \int_{0}^{\infty} d\eta \ 2\pi^2   \left[ \frac{(u')^2}{1-u^2} 
+ V_{m,n}(\eta) (1-u^2) \right] 
  \nonumber\\
 &+  C_4 a^{-1} \int_{0}^{\infty} d\eta \ 2\pi^2  (u')^2 \cosh \eta \sinh \eta V_{m,n}(\eta) ,
\end{align}
where 
$n_3 = u(\eta) := \frac{f^2(\eta)-1}{f^2(\eta)+1}$, 
$u':=\frac{du}{d\eta}$, 
and
$
 V_{m,n}(\eta) :=   m^2+n^2/\sinh^2 \eta   .
$
  We obtain the knot soliton solution $\{ \vec{n}(\vec{x}) \}$ by varying the energy functional $E=\Lambda_{\rm F}^2 a E^{(2)} + C_4 a^{-1} E^{(4)}$ with respect to $u$ and $a$,  
where we impose the boundary condition
$f(\eta=0) = \infty$, i.e., the north-pole $\vec{n}=(0,0,1)$ at $z$-axis  
and  
$f(\eta=\infty) = 0$, i.e., the south-pole $\vec{n}=(0,0,-1)$ at the circle $C:=\{ x^2+y^2=a^2, z=0  \}$.

The scaling relation corresponding to (\ref{E-scaling}) is easily derived 
in the toroidal ansatz: 
$
 E =\Lambda_{\rm F} C_4^{1/2} [ \tilde{a}E^{(2)} + \tilde{a}^{-1} E^{(4)}]
$ 
using  a dimensionless parameter $\tilde{a}$ defined by 
\begin{align}
 a = \Lambda_{\rm F}^{-1} C_4^{1/2} \tilde{a} .
\label{a-scaling}
\end{align} 
The variation of $E$ with respect to $\tilde{a}$ yields  
$
 \tilde{a} = \sqrt{E^{(4)}/E^{(2)}}
$
which determines the radius 
$ 
 a= \Lambda_{\rm F}^{-1} C_4^{1/2} \tilde{a}
  = \Lambda_{\rm F}^{-1} C_4^{1/2} \sqrt{E^{(4)}/E^{(2)}}
$
of the circle $C$, i.e., size of the knot soliton.  
This relation is substituted back into the energy functional to obtain
$
 E=2\Lambda_{\rm F} C_4^{1/2} \sqrt{E^{(2)}E^{(4)}}
$
which is further varied with respect to $u$. 
The energy has the upper bound 
$
 E \le \Lambda_{\rm F} C_4^{1/2} (E^{(2)}+E^{(4)})
$.

\begin{figure}[htbp]
\begin{center}
\includegraphics[height=5cm]{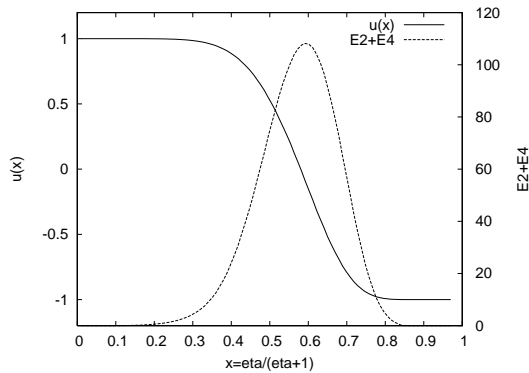}
\caption{\small 
The knot soliton solution $u$  at $\Lambda_{\rm F} =1$ and $C_4=1$
for $m=1, n=2; Q_H=mn=2$.
(Left vertical axis) $n_3=u$ vs. $x=\eta/(1+\eta)$, 
(Right  vertical axis) energy density $\varepsilon$ vs. $x$ where 
$\int_{0}^{\infty}d\eta \varepsilon(\eta)=E^{(2)}+E^{(4)}$.
}
\label{fig:knot-solution}
\end{center}
\end{figure}

In Fig.~\ref{fig:knot-solution}, we plot the numerical solution of the variable  
$n_3 = u(\eta) := \frac{f^2(\eta)-1}{f^2(\eta)+1}$ for a knot soliton 
satisfying boundary conditions  $u(0)=1$ and $u(\infty)=-1$.
The energy density $\varepsilon(\eta)$  as a function of $\eta$  is concentrated on a finite width which represents the location (thickness) of the knot.
The rest mass $M$ of the knot soliton  is equal to the energy of the static knot soliton.
Our numerical result (Table~1 below) shows e.g., 
$
 M_{*}=465
$ 
for $m=1, n=2; Q_H=mn=2$, 
which is consistent with the previous results \cite{HS98,GH97,SST06}. 


\section{Quantization of knot soliton}


Second, 
we perform the collective coordinate quantization  of the knot soliton solution for the Faddeev model. 
The collective coordinate quantization for the Faddeev model has already been investigated in \cite{KS05,Su}. 
However, the author of \cite{Su} has used a different background configuration from ours. The authors of \cite{KS05} have given more general and mathematical framework for the quantization of the knot soliton than the collective coordinate quantization in the Faddeev model.
Moreover, they have discussed the collective coordinate quantization too. However,  the necessary expressions and the details of calculations to reproduce their results were not written in the paper \cite{KS05}, which disables us from utilizing their results. 
In this paper, we have worked out all the steps of collective coordinate quantization to give explicit expressions and numerical data, following the way similar to \cite{BC88}.  
Furthermore, we give an argument for the relationship between the Faddeev model and the original Yang-Mills theory, although the paper \cite{KS05} aimed at a  physical application due to the fermionic quantization of knot solitons which is quite different from ours based on the bosonic quantization. 

We observe that for a generic configuration $\vec{n}(\vec{x})$, there exists a nine-parameter set of configurations, all degenerate in energy (zero modes), obtained from  $\vec{n}(\vec{x})$ by some combination of translation ($\vec{X}$), spacial rotation ($B$), and isospin rotation ($A$): 
\begin{align}
 \bm{n}(\vec{x}) \rightarrow \bm{n}'(\vec{x}) = A \bm{n}(R(B)  (\vec{x}-\vec{X})) A^\dagger, \quad A, B \in SU(2) 
\end{align}
where we have used the Lie-algebra $su(2)$  valued field:
$
 \bm{n}  := n_A T_A, \quad T_A = \frac12 \sigma_A ,
$
and
the spacial rotation is represented by  orthogonal matrices 
$R_{jk}(B)=\frac12 {\rm tr}(\sigma_j B \sigma_k B^\dagger)$. 
The minimal or semi-classical quantization called the collective coordinate quantization proceeds by promoting the parameters $\vec{X}, A, B$ to the time-dependent dynamical variables $\vec{X}(t), A(t), B(t)$ to remove the classical degeneracy of the static configuration.  
Then the dynamical ansatz adopted in collective coordinate quantization
 is 
\begin{align}
 \bm{n}(\vec{x}) \rightarrow \bm{n}'(\vec{x},t) = A(t) \bm{n}(R(B(t))  (\vec{x}-\vec{X}(t))) A^\dagger(t), \quad A, B \in SU(2) ,
\end{align}
where the prime serves  distinguishing the dynamical variable $\bm{n}'(\vec{x},t)$ from the static background field $\bm{n}(\vec{x})$.
The translational degrees of freedom $\vec{X}(t)$ can be ignored if the knot soliton is quantized in its rest frame. 
By inserting this ansatz into the Faddeev Lagrangian density $\mathscr{L}_{\rm F}$ and integrating over the three-dimensional space, the Lagrangian  in the rest frame is determined to be%
\footnote{ 
Without the restriction to the rest frame, the kinetic energy $T$ has extra terms of the type $a_j Y_{jk} \dot{X}_k$ and $b_jZ_{jk} \dot{X}_k$, in addition to 
$\frac12 \dot{X}_j M_{jk} \dot{X}_k$. 
}
\begin{align}
 L_{\rm F} =& \int d^3x \mathscr{L}_{\rm F}[\bm{n}', \partial \bm{n}'] = T-M ,
\nonumber\\
 T =& 
 \frac12 a_j U_{jk} a_k + a_j W_{jk} b_k + \frac12 b_j V_{jk} b_k ,
\end{align}
where the variables $a_j, b_k$ are defined by
\begin{align}
   a_j := -i {\rm tr}(\sigma_j A^\dagger \dot{A}) \quad  \text{isospin~rotation},
\quad 
   b_j :=  i {\rm tr}(\sigma_j \dot{B} B^\dagger ) \quad \text{spacial~rotation} , 
   \label{def}
\end{align}
and the inertial tensors are given by
\begin{align}
 U_{jk} =& \int d^3x \left\{ \Lambda_{\rm F}^2 (\delta_{jk}-n_j n_k ) + C_4 \partial_\ell n_j  \partial_\ell n_k  \right\} 
 \nonumber\\
   =&  \int d^3x \left\{ -2 \Lambda_{\rm F}^2 {\rm tr}([T_j, \bm{n}][T_k, \bm{n}]) + 2 C_4 {\rm tr}([[T_j, \bm{n}], \partial_\ell \bm{n}][[T_k, \bm{n}], \partial_\ell \bm{n}]) \right\} ,
  \nonumber\\
 W_{jk} =&  - \int d^3x \left\{ \Lambda_{\rm F}^2 (\vec{n} \times iL_{k} \vec{n})_{j} + C_4 \partial_\ell n_j \vec{n} \cdot (iL_{k} \vec{n} \times \partial_\ell \vec{n}) \right\} ,
 \nonumber\\
 V_{jk} =&   \int d^3x \left\{ \Lambda_{\rm F}^2 (iL_{j} \vec{n}) \cdot (iL_{k} \vec{n}) + C_4 (iL_{j} \vec{n} \times \partial_\ell \vec{n}) \cdot (iL_{k} \vec{n} \times \partial_\ell \vec{n})  \right\} ,
\end{align}
with the angular momentum operator 
$\vec{L}=(L_j), L_j := \epsilon_{jk\ell}x_k p_\ell = -i \epsilon_{jk\ell} x_k \partial_\ell$.
Here, $W_{jk}$ and $V_{jk}$ are obtained from $U_{jk}$ by the replacement: 
$W_{jk}=U_{jk}[[T_k, \bm{n}] \rightarrow L_k \bm{n}]$
and 
$V_{jk}=W_{jk}[[T_j, \bm{n}] \rightarrow L_j \bm{n}]
= U_{jk}[[T_j, \bm{n}] \rightarrow L_j \bm{n}; [T_k, \bm{n}] \rightarrow L_k \bm{n}]$.

The scaling relation  for the inertial tensor reads, e.g., 
\begin{align}
 V_{jk}[\vec{n}(\vec{x})]_{\Lambda_{\rm F},C_4}  =\Lambda_{\rm F}^{-1} C_4^{3/2} V_{jk}[\vec{n}(\Lambda_{\rm F}^{-1} C_4^{1/2}\vec{x})]_{1,1} .
\end{align} 
The inertial tensors under the toroidal ansatz have the following properties reflecting the symmetries of the classical background configurations considered in the above.
\begin{enumerate}
\item[i)]
$U_{jk}, W_{jk}$ and $V_{jk}$ are diagonals and all off-diagonal components vanish except for $W_{21}=nW_{12} \ne 0$ ($n = \pm 1$).

\item[ii)]
a) $U_{11}=U_{22}$, $V_{11}=V_{22}$,

b) $V_{33}=nW_{33}=n^2U_{33}$ .

\item[iii)]
$W_{11}=0= W_{22}$ .

\end{enumerate}
The property i) follows from the fact that the off-diagonal components $j\ne k$ vanish  after the angular integration over $\varphi$ by its periodicity. 
The property ii-a) for the perpendicular components is a direct consequence of the cylindrical symmetry, while ii-b) for the parallel ones comes from the identify which holds for the toroidal ansatz:
\begin{align}
 L_3 \bm{n} = - i \frac{\partial}{\partial \varphi} \bm{n} = - n [T_3, \bm{n}] .
\end{align}
The property iii) 
follows also from explicit calculations.   
The angular integration over $\xi$ is necessary to show 
$W_{11}=0= W_{22}$ for all integers $n$ and $m$, since the angular integration over $\varphi$ alone leads to $W_{11}=0= W_{22}$ ($n\ne \pm 1$) and   $W_{11}=-nW_{22} \ne 0$ ($n = \pm 1$).  This is because the integrand is an odd function of $\xi$ where  $\xi \rightarrow -\xi$ corresponds to the reflection with respect to the $z=0$ plane, i.e., $x \rightarrow x$, $y \rightarrow  y$, $z \rightarrow -z$.

After lengthy calculations, we obtain the explicit form:
\begin{align}
V_{11} =& V_{11}^a + V_{11}^b ,
\nonumber\\
 V_{11}^a =& \int_{0}^{\infty} d\eta  \
 \Lambda_{\rm F}^2 a^3 \pi^2    \Biggr\{ 
 \coth^2\eta  \frac{(u')^2}{1-u^2} 
+ 
 \left[ m^2 ( 3\coth^2 \eta -2)  + \frac{n^2}{\sinh^4\eta}   \right] (1-u^2) \Biggr\} ,
 \nonumber\\
 V_{11}^b =& 
 \int_{0}^{\infty} d\eta    \ 
 C_4 a 2 \pi^2    \Big[ 
 m^2  \cosh\eta \sinh\eta  
 + 
   n^2  (1+\cosh^2\eta) (\coth \eta-1)   \Big] (u')^2 ,
\label{V_11}
\end{align}
\begin{align}
  U_{33} = U_{33}^a + U_{33}^b
 = \int_{0}^{\infty} d\eta \Big\{ \Lambda_{\rm F}^2 a^3 2\pi^2 \frac{3\coth^2\eta-1}{\sinh^2\eta}(1-u^2) 
 + C_4 a 4\pi^2  (u')^2 \Big\} ,
 \label{U_33}
\end{align}
\begin{align}
U_{11} =  U_{11}^a + U_{11}^b , \quad
U_{11}^a =& \int_{0}^{\infty} d\eta \   \Lambda_{\rm F}^2 a^3  \pi^2 \frac{3\coth^2\eta-1}{\sinh^2\eta}(1+u^2) ,
\nonumber\\
 U_{11}^b =& \int_{0}^{\infty} d\eta \  C_4 a 2\pi^2    \left[ \frac{u^2 (u')^2}{1-u^2} + (1-u^2) V_{m,n}(\eta) \right]  .
 \label{U_11}
\end{align}
\begin{align}
W_{12} =&  W_{12}^a + W_{12}^b ,  
\nonumber\\
W_{12}^a =& - \Lambda_{\rm F}^2a^3 \pi^2 \Biggr[ m
    \int_{0}^{\infty} d\eta
    \Bigl\{
    \cosh(m\eta)-\sinh(|m|\eta)
    \Bigr\}
    (\coth\eta+|m|)
    \nonumber\\
 &\qquad\qquad\qquad\qquad
    \times
    \left(
    \coth\eta
     \frac{u^\prime}{\sqrt{1-u^2}}
    -\frac{u\sqrt{1-u^2}}{\sinh^2\eta}
    \right)
    \nonumber\\
 &\quad\quad\quad\quad
    +    m
     \int_{0}^{\infty} d\eta
     \biggl\{
   \frac{3\coth\eta}{\sinh^4\eta}
   +|m| \frac{3\coth^2\eta-2}{\sinh^2\eta}
   +m^2\frac{\coth\eta}{\sinh^2\eta}
     \biggr\}
   \nonumber\\
 &\qquad\qquad\qquad\qquad
     \times
     \Bigl\{
     \cosh(m\eta)-\sinh(|m|\eta)
     \Bigr\}
      \sinh^2\eta
     u\sqrt{1-u^2} 
     \Biggr] ,
\nonumber\\
 W_{12}^b =& -
     C_4 a 2\pi^2 \Biggr[ 
    \frac{m}{|m|}
    \int_{0}^{\infty} d\eta
    \Bigl\{
    \cosh(m\eta)-\sinh(|m|\eta)
    \Bigr\}
    \nonumber\\
 &\qquad\qquad\qquad\qquad
    \times
    \left\{
    \sinh\eta
    \cosh\eta
    V_{m,n}(\eta)
    u^\prime\sqrt{1-u^2}
    -\frac{(u^\prime)^2 u}{\sqrt{1-u^2}}
    \right\}
   \nonumber\\
 &\quad\quad\quad\quad
   +    m
    \int_{0}^{\infty} d\eta
    \Bigl\{
    \cosh(m\eta)-\sinh(|m|\eta)
    \Bigr\}
    \sinh\eta
    \cosh\eta
    \frac{(u^\prime)^2 u}{\sqrt{1-u^2}} 
    \Biggr] .
 \label{W_12}
\end{align}

The  conjugate momenta corresponding to $X_j$, $A$ and $B$ are 
$ 
 P_j := \frac{\partial L_{\rm F}}{\partial \dot{X}_j},
 \quad  K_j := \frac{\partial L_{\rm F}}{\partial a_j}, \quad L_j := \frac{\partial L_{\rm F}}{\partial b_j} ,
$ 
since the variables $a_j, b_k$ defined by (\ref{def}) are regarded as generalized velocities in the terminology of analytical dynamics.  
Here $\vec{L}$ and $\vec{K}$ are body-fixed spin and body-fixed isospin angular momenta.  (Isospin $\vec{K}$ becomes the angular momentum operator acting on the target space.)  
The Hamiltonian defined through the Legendre transform,  
$
 H_{\rm F}  := P_j \dot{X}_j  + K_j a_j + L_j b_j - L_{\rm F}  ,
$
is expressed in terms of  $K_j, L_j, P_j$ after eliminating $\dot{X}_j, a_j, b_k$. 
Here it should be remarked that there is a primary constraint:
\begin{align}
 L_3 = -n K_3 ,
\end{align}
which yields a selection rule for possible glueball spectrum below.

The Hamiltonian in the rest frame $\dot{X}_j=0$ reads
\begin{align}
 H_{\rm F}  =& M  + \frac{1}{2} \frac{1}{U_{11}-\frac{W_{12}^2}{V_{11}}}(K_{1}^2+K_{2}^2) 
 + \frac{1}{2} \frac{1}{V_{11}-\frac{W_{12}^2}{U_{11}}} (L_{1}^2+L_{2}^2) 
 \nonumber\\
 & + \frac{1}{W_{12}-\frac{U_{11}V_{11}}{W_{12}}} K_{1}L_{2}
 + \frac{1}{W_{21}-\frac{U_{11}V_{11}}{W_{21}}} K_{2}L_{1} 
  + \frac{1}{2U_{33}} K_{3}^2  .
\end{align}
Note that $U_{11}=U_{22}$ diverges $U_{11}=U_{22}=O(V) \rightarrow \infty$ as the volume $V$ of the three-dimensional space.  Therefore the Hamiltonian reduces to 
\begin{align}
 H_{\rm F} = 
  M +  \frac{1}{2V_{11}} (L_{1}^2+L_{2}^2) + \frac{1}{2U_{33}} K_{3}^2 
 =  M +   \frac{1}{2V_{11}} \vec{L}^2 + \frac{1}{2} \left( \frac{1}{U_{33}} - \frac{n^2}{V_{11}} \right) K_{3}^2 .
\label{Hamiltonian}
\end{align}
The Hamiltonian has the same form as the symmetrical top \cite{Rose}. 
Thus we obtain the energy eigenvalue in the rest frame of the knot soliton:
\begin{align}
  E  = M    + \frac{1}{2V_{11}} J(J+1) + \frac{1}{2} \left( \frac{1}{U_{33}} - \frac{n^2}{V_{11}} \right) K_{3}^2 ,
\end{align}
where we have used the relationship $\vec{L}^2=\vec{J}^2$ between the body-fixed spin angular momentum $\vec{L}$ and the space-fixed (coordinate-fixed) spin angular momentum $\vec{J}$, since they are related to each other by spacial rotations:  
$J_j=-R_{jk}(B)^T L_k$. 
For rotations in the target space, only rotations around the third axis are compatible with the boundary conditions and hence we have a relationship
$K_3^2=I_3^2$
between the body-fixed isospin angular momentum $\vec{K}$ and the space-fixed  (coordinate-fixed) isospin angular momentum $\vec{I}$ which are related to each other by rotations:  
$I_j=-R_{jk}(A)K_k$.

Thus the  energy of the  knot soliton is modified after quantization into 
\begin{align}
  E  = \Lambda_{\rm F} C_4^{1/2} M_{*}    + \Lambda_{\rm F} C_4^{-3/2} \left[ \frac{1}{2V_{11}^{*}} J(J+1) + \frac{1}{2} \left( \frac{1}{U_{33}^{*}} - \frac{n^2}{V_{11}^{*}} \right) K_{3}^2 \right] ,
\end{align}
where  we have used the scaling argument for inertial tensors (\ref{V_11})-(\ref{U_11}) and $a$ must be replaced by $\tilde{a}$ in $V_{11}^{*}$ and $U_{33}^{*}$, which follows from the substitution of (\ref{a-scaling}).

\section{Symmetries and constraints}

Every symmetry of a classical configuration induces a loop in configuration space. After quantization, these loops give rise to constraints on the wavefunction.
In the followings, we study symmetries given by a rotation by $\alpha$ in space and by a rotation $\beta$  in target space (isorotation). 
For example, one can consider a special path associated with the symmetries parameterized by   $\alpha$ angle about $k$ axis  in space and $\beta$ angle about $j$ axis in target space: 
\begin{equation}
 A(\beta) := e^{i\beta K_j } A e^{-i\beta K_j } = A e^{i\beta \sigma_j/2} , \quad 
 B(\alpha) := e^{i\alpha L_k } B e^{-i\alpha L_k } = e^{-i\alpha \sigma_j/2} B ,
 \label{path-angle}
\end{equation}
which follow from the commutation relations
\begin{equation}
 [K_j, A] = \frac12 A \sigma_j, \quad
 [L_k, B] = - \frac12 \sigma_k B .
\end{equation}
Then the end-point operator is given by
\begin{equation}
 L(\alpha, \beta) = e^{i\beta K_j } e^{i\alpha L_k } .
\end{equation}
If such rotations become symmetries inducing loops in configuration space, the {\it Finkelstein-Rubinstein (FR)  constraint} \cite{FR68} must be imposed on the wavefunction $\psi$ after quantization:
\begin{equation}
 \exp (i \beta \vec{N} \cdot \vec{K} ) \exp (i\alpha \vec{e} \cdot \vec{L})  \psi = \chi_{FR} \psi ,
\end{equation}
where $\vec{e}$ is the direction of the rotation axis in space, $\vec{N}$ is the rotation axis in target space (isospace), 
$\vec{L}$ and $\vec{K}$ are the angular momentum operators in space and target space, respectively.

A scalar field theory can be quantized by considering the wave function on configuration space. This gives a purely bosonic theory. If the fundamental group of the configuration space is $\mathbb{Z}_2$, one can also consider wave functions on the covering space of configuration space. This gives rise to a quantum theory which contains fermions, provided so-called Finkelstein-Rubinstein constraints are introduced.

In the quantization of solitons, therefore, two choices are possible; fermionic quantization or bosonic quantization. 
In the {\it fermionic quantization}, the FR phase $\chi_{FR}$ is given by
\begin{equation}
 \chi_{FR} = 
 \begin{cases}
 +1 & \text{if the induced closed loop is contractible} \\
 -1 & \text{otherwise (non-contractible)} 
 \end{cases} .
\end{equation}
Whereas the {\it bosonic quantization} adopts the  FR phase
\begin{equation}
 \chi_{FR} = +1 , 
\end{equation}
irrespective of whether the induced closed loop is contractible or not.

There are several arguments why the Skyrme model has to be quantized fermionically. 
For a rotation by $2\pi$, the following facts are known  
for the Skyrme model.
Williams \cite{Williams70} and Giulini \cite{Giulini93} have  shown that the  configuration space admits spinorial states, i.e., non-contractible loop with $\chi_{FR}=-1$ in the sectors of odd topological charge (winding number or baryon number) $B$, and no spinorial states, namely, only contractible loops with $\chi_{FR}=+1$  are induced in the sectors of even topological charge $B$.
Finkelstein and Rubinstein \cite{FR68} have shown that a $2\pi$ rotation of a Skyrmion of charge $B$ is homotopic to an exchange of two Skyrmions of charge $B$. This implies that an exchange of two identical Skyrmions gives rise to $\chi_{FR}=-1$ if and only if their topological charge $B$ is odd, or $\chi_{FR}=+1$ if and only if $B$ is even. 
The Skyrme model allows for fermionic quantization with half-odd angular momentum in the odd $B$ sectors, and bosonic quantization with integer angular momentum only in the even $B$ sectors. This is the topological spin-statistics theorem for the Skyrmion \cite{SSS}. 
Krusch \cite{Krusch03} has shown that a $2\pi$ isorotation of a Skyrmion also gives rise to $\chi_{FR}=-1$ if and only if $B$ is odd.

 For Skrymions which are well-approximated by the rational map ansatz (rational map Skyrmion), a simple formula for the FR phase in the fermionic quantization has been given \cite{Krusch03},
$\chi_{FR}=(-1)^{N}$, $N=B(B\alpha-\beta)/2\pi$, and the formula was generalized to more complicated situations in \cite{Krusch05}.

Here it is instructive to recall the consideration done by Braaten and Carson \cite{BC88} for the $B=2$ Skyrmion under the product Ansatz.
For the $B=2$ Skyrmion, the paths induced by $\pi$ rotations about coordinate axes are closed in the sense of returning the transformed field at $\theta=\pi$ to its original value at $\theta=0$:
$U(\vec{r},t)=\exp (i \beta \vec{N} \cdot \vec{K} ) \exp (i\alpha \vec{e} \cdot \vec{L}) U(\vec{r},t) \exp (-i\alpha \vec{e} \cdot \vec{L}) \exp (-i \beta \vec{N} \cdot \vec{K} )$ for the Skyrme field $U(\vec{r},t)$. 
Consequently, acting the end-point operator 
$ \exp (i \beta \vec{N} \cdot \vec{K} ) \exp (i\alpha \vec{e} \cdot \vec{L}) $
on the wavefunction gives rise to a phase $\chi_{FR}=+1$ or $-1$ depending on whether the associated path is contractible or not, respectively. 
The analysis of \cite{BC88} shows that $\pi$ rotations about $x$ and $y$ axes with end-point operators $e^{i\pi K_1}e^{i\pi L_1}$ and $e^{i\pi K_1}e^{i\pi L_2}$ correspond to non-contractible paths leading to $\chi_{FR}=-1$, while $\pi$ rotation about the $z$ axis with the end-point operator $e^{i\pi L_3}$ corresponds to a contractible path leading to $\chi_{FR}=+1$. 

In this paper, we intend to use the Faddeev model for describing effectively glueballs which are bosons with integer spins to be identified with composite particles made of spin-one gluons.  In this paper, therefore, we choose the bosonic (collective coordinate) quantization of knot solitons in the Faddeev model.  
(This choice should be compared with the fermionic quantization chosen in Krusch and Speight \cite{KS05}.) 
In view of this, we study the symmetries and the corresponding constraints on the wavefunction to be imposed after quantization. 
In the Faddeev model under the simple toroidal ansatz, the minus sign in (\ref{x}) or (\ref{y}) can not be expressed by using $A(\beta)$ and $B(\alpha)$ in the manner described in (\ref{path-angle}). 
Rather, it denotes a reflection in the target space. 
In any case, therefore, the spacial rotations by $\pi$ radians about coordinate axes $x$ and $y$ do not represent closed loops in the configuration space of the color field, while the spacial rotation by $\pi$ radians about $z$ axis (\ref{z}) represents a closed loop, as a special case of the continuous axial symmetry $\varphi \rightarrow \varphi + \gamma$, i.e., $\gamma =\pi$. 
Therefore, to obtain  closed loops induced by rotations about  coordinate axes $x$ and $y$, the rotation angle must be at least $2\pi$ radians.

In quantum system, the operators 
$\vec{\hat{J}}^2=\vec{\hat{L}}^2, \hat{L}_3, \hat{J}_3, \hat{I}_3$ and $\hat{K}_3$ form a set of commuting observables and the eigenfunction $\psi$ can be labelled by the quantum numbers 
$L, L_3, J_3, I_3$ and $K_3$. 
We consider states with given $J$ and $I_3$, namely, $L=J$ and $K_3=\pm I_3$ and $L_3=-nK_3$.
Then the eigenfunction for the above Hamiltonian is given up to a normalization constant 
by $\psi=|I,I_3,K_3 \rangle \times |L,J_3,L_3 \rangle$ 
where $|I,I_3,K_3 \rangle$ corresponds to 
a finite rotation in the target space, i.e, isorotation $A \in SU(2)/Z_2=SO(3)$,
and $|J,J_3,L_3 \rangle$ 
to a finite spacial rotation $B \in SO(3)$.

Therefore, reflecting the transformation law (\ref{x}), (\ref{y}) and  (\ref{z}) of the classical configuration for the $\pi$ rotations about $x$, $y$ and  $z$ axes, 
 the end-point operators representing closed loops, namely,  the $2\pi$ rotations about $x$ and $y$ axes and the $\pi$ rotation about $z$ axis  are constructed to act on the wavefunction 
$\psi=|I,I_3,K_3 \rangle \times |L,J_3,L_3 \rangle$ in the following way: 
for $n=$ even ($L_3=-nK_3=$ integer): 
\begin{subequations}
\begin{align}
 e^{2\pi i \hat{K}_2} e^{2\pi i \hat{L}_1}  \psi =& (-1)^{2L+2I} \psi = \chi_{\rm FR} \psi,
 \\
 e^{2\pi i \hat{K}_2} e^{2\pi i \hat{L}_2}  \psi =& (-1)^{2L+2I} \psi = \chi_{\rm FR} \psi,
 \\
  e^{\pi i \hat{L}_3}  \psi =& (-1)^{L_3} \psi = \chi_{\rm FR} \psi.
\end{align}
\label{FR1}
\end{subequations}
While, for $n=$ odd ($L_3+K_3=(1-n)K_3=$ integer), we have
\begin{subequations}
\begin{align}
 e^{2\pi i \hat{K}_2} e^{2\pi i \hat{L}_1}  \psi =& (-1)^{2L+2I} \psi = \chi_{\rm FR} \psi,
 \\
 e^{2\pi i \hat{K}_1} e^{2\pi i \hat{L}_2}  \psi =& (-1)^{2L+2I} \psi = \chi_{\rm FR} \psi,
 \\
  e^{\pi i \hat{K}_3}  e^{\pi i \hat{L}_3}  \psi =& (-1)^{L_3+K_3} \psi= (-1)^{(1-n)K_3} \psi= \chi_{\rm FR} \psi  .
\end{align}
\label{FR2}
\end{subequations}
Here we have used the properties of the Wigner D-function \cite{Rose}:
$
\langle J,M'|e^{i\pi \hat{J}_y}|J,M \rangle =\delta_{M',-M}(-1)^{J+M}
$
and
$
\langle J,M'|e^{i\pi \hat{J}_x}|J,M \rangle =\delta_{M',-M}(-1)^{J+M} e^{-i\pi M} 
$,
leading to 
$
 e^{i\pi \hat{J}_x}|J,M \rangle = (-1)^{J+M}e^{-i\pi M} |J,-M \rangle
$
and
$
 e^{i\pi \hat{J}_y}|J,M \rangle =(-1)^{J+M}|J,-M \rangle
$,
in addition to  the relation
$
 e^{i\pi \hat{J}_z}|J,M \rangle =e^{i\pi M}|J,M \rangle =(-1)^{M}|J,M \rangle
$. 
They are applied to our case to obtain the formulae:
\begin{align}
 e^{i\pi \hat{L}_1}|L,J_3,L_3 \rangle =&  (-1)^{L+L_3}e^{-i\pi L_3} |L,J_3,-L_3 \rangle ,
\nonumber\\
 e^{i\pi \hat{L}_2} |L,J_3,L_3 \rangle =& (-1)^{L+L_3}|L,J_3,-L_3 \rangle ,
\nonumber\\
 e^{i\pi \hat{L}_3} |L,J_3,L_3 \rangle =& (-1)^{L_3}|L,J_3,L_3 \rangle ,
\end{align}
and
\begin{align}
 e^{i\pi \hat{K}_1}|I,I_3,K_3 \rangle =&  (-1)^{I+K_3}e^{-i\pi L_3} |I,I_3,-K_3 \rangle ,
\nonumber\\
 e^{i\pi \hat{K}_2} |I,I_3,K_3 \rangle =& (-1)^{I+K_3}|I,I_3,-K_3 \rangle ,
\nonumber\\
 e^{i\pi \hat{K}_3} |I,I_3,K_3 \rangle =& (-1)^{K_3}|I,I_3,K_3 \rangle .
\end{align}

Besides rotations considered in the above, closed loops are indeed obtained by $2\pi$ rotations in space and the target space, and our restriction for the FR constraint  for these loops in the bosonicc quantization $\chi_{\rm FR}=+1$ leads to 
\begin{equation}
e^{2\pi i K_3} \psi =(-1)^{2K_3} \psi = \chi_{\rm FR} \psi , \quad  
e^{2\pi i L_3} \psi =(-1)^{2L_3} \psi= \chi_{\rm FR}  \psi , 
\end{equation}
which implies that 
$L_3$ and $K_3$ must be integers. 

Moreover, we examine the restriction on quantum numbers coming from the FR constraints  for the  symmetries (\ref{FR1}) and (\ref{FR2}) in the bosonic quantization $\chi_{\rm FR}=+1$. A  condition, $I+L=$ integer, must be satisfied for any $n$.
For $n=$ even, $L_3=-nK_3$= even integer and hence $L=J$ must be integer (excluding half-odd integer) yielding that $I$ is also an integer. 
For $n=$ odd,  we have also a condition, $I+L=$ integer.  In this case, we can conclude only that $L_3$ and $K_3$ are integers. Hence $L=J$ is also an integer yielding that $I$ is also an integer. 
Thus, both $I$ and $J$ are integers. 
These results are reasonable for our purposes of identifying the Hopf solitons with bosons and can be a self-consistency check for choosing the bosonic quantization.
Further symmetries may impose additional constraints for quantum numbers, leading to further selection rules for possible spectrum of glueballs.

\begin{table} 
\small
\begin{center}
 \begin{tabular}{|c|c|c|}
  \hline
  \multicolumn{3}{|c|}{SU(2) Lattice gauge theory}  \\ \hline
   $J^{PC}$  & $M_G/\sqrt{\sigma}$ & $M_G/M_{0^{++}}$ \\ \hline
   $0^{++}$  & 3.74 $\pm$ 0.12 & 1 \\
   $2^{++}$  & 5.62 $\pm$ 0.26 & 1.46 $\pm$ 0.094 \\
   $0^{-+}$  & 6.53 $\pm$ 0.56 & 1.78 $\pm$ 0.24 \\
   $2^{-+}$  & 7.46 $\pm$ 0.50 & 2.03 $\pm$ 0.20 \\
   $1^{++}$  & 10.2 $\pm$ 0.5 & 2.75 $\pm$ 0.15 \\
   $1^{-+}$  & [10.4 $\pm$ 0.7] & [3.03 $\pm$ 0.31] \\
   $3^{++}$  & 9.0 $\pm$ 0.7 & 2.46 $\pm$ 0.23 \\
   $3^{-+}$  & [9.8 $\pm$ 1.4] & [2.91 $\pm$ 0.47] \\ \hline
 \end{tabular}
\caption{
Glueball masses, (right column) in units of the lightest scalar glueball mass and (left column) of the string tension, in the continuum limit of SU(2) Yang-Mills theory on a lattice \cite{Teper98}.
Values in brackets have been obtained by extrapolating from only two lattice values and so should be treated with caution. 
}
\label{table:glueball-mass}
\end{center}
\end{table}

\section{\large  Glueball mass, gluon mass and vacuum condensate}

As in the previous section, we restrict the following considerations to bosonic quantization with the trivial FR phase $\chi_{FR}=+1$. 
Then the eigenfunction has  the simplified form:
\begin{align}
  & \psi = |L,L_3,K_3 \rangle ,
  \nonumber\\
 & J= 0, 1, 2, \dots, \quad |K_3| = 0, 1, \cdots, J, \quad 
  L_3= 0, \pm 1, \cdots, \pm J ,
\end{align}
where 
we have suppressed  $J_3$ and $I_3$, since we find no constraints on the values of $J_3$ and $I_3$ according to \cite{KS05}. 
This wavefunction is identified with the Wigner D-function \cite{Rose} 
$
 \psi^J_{L_3K_3} = C D^{(J)}_{L_3K_3}(\alpha, \beta, \gamma)
$
for a finite rotation with the Euler angles  $\alpha, \beta, \gamma$ where $C$ is a normalization constant $C=\sqrt{(2J+1)/(8\pi^2)}$, which is known as the eigenfunction of the above Hamiltonian (\ref{Hamiltonian}) of a symmetrical top.

In the followings, we consider only  quantum states with Hopf charge $|Q_H|=1, 2$ as candidates for  glueball states constructed from gluons with spin-one. 
This is because the simplified toroidal ansatz adopted in this paper is valid only for $|Q_H|=1, 2$ according to numerical calculations \cite{HS98}. 
See Table.\ref{table:glueball-mass} for results of numerical simulations for glueballs on a lattice. 
For $|Q_H|=1$, we have a case $(n,m)=(1,1)$.
For $|Q_H|=2$, our numerical calculations for $Q_H=2$ show that the classical energy $M_{*}$ (the rest mass of the knot soliton) for $(n,m)=(2,1)$ is lower than the value for $(n,m)=(1,2)$, see Table~\ref{table:numerical-results}.  

For $m=n$, i.e., $Q_H=m^2=1,4,9, \cdots$, the Vakulenko-Kapitanskii lower-bound on the energy \cite{VK79} by the Hopf index, 
$E \ge  {\rm const.} |Q_H|^{3/4}$,  is saturated \cite{LY04} by the classical background configuration, although the generic Hopf soliton is not of the BPS type.  In this case, the lowest $0^{++}$ glueball mass is expressed as
\begin{align}
  M_{0^{++}} = {\rm const.} \Lambda_{\rm F} C_4^{1/2}  |Q_H|^{3/4}
  \cong {\rm const.} g^{-1} \sqrt{\left< -\mathbb{X}_\mu^2 \right>}|Q_H|^{3/4} ,
\end{align} 
by the energy of 
the Hopfion with non-zero Hopf index $Q_H$.  
The non-perturbative dependence $g^{-1}$ of the mass on the coupling constant is characteristic of the soliton.

For definiteness, we restrict the following consideration to the case $(n,m)=(2,1)$, since the other cases can be treated exactly in the same way. 
We adopt a simple identification as inputs: 
\begin{align}
 0^{++} \leftrightarrow |L,L_3,K_3 \rangle  = |0,0,0 \rangle    \ (m=1, n=2; Q_H =2) ,
\nonumber\\
 2^{++}  \leftrightarrow |L,L_3,K_3 \rangle  = |2,0,0 \rangle \ (m=1, n=2; Q_H =2) .
\end{align}
Note that Parity is given by 
$P=(-1)^J$ for $K_3=0$ 
and the wavefunction reads  
$
D^{(J)}_{00}(\varphi,\theta,-\varphi) = d^{J}_{00}(\theta) = P_J(\cos \theta)  
$ 
with no $\varphi$-dependence for $K_3=0=L_3$ .
Under this identification, the lowest two glueball masses are given by
\begin{align}
 M_{0^{++}} = \Lambda_{\rm F} C_4^{1/2} M_{*} ,
 \quad 
 M_{2^{++}} =  \Lambda_{\rm F} C_4^{1/2} M_{*} +  3 \Lambda_{\rm F} C_4^{-3/2}/V_{11}^{*},
\end{align}
which yield  the mass ratio
$
  M_{2^{++}}/M_{0^{++}} = 1+  3C_4^{-2}/(M_{*}V_{11}^{*}) .
$

\begin{table} 
\small
\begin{center}
 \begin{tabular}{|c|c|c|c|c|c|c|}
  \hline
  \multicolumn{7}{|c|}{SU(2) Faddeev model}  \\ \hline
   $Q_H$  &  $(n,m)$ & $M_{*}$   & $V_{11}^{*}$  & $U_{33}^{*}$ & $W_{12}^{*}$ & $\tilde{a}$ \\ \hline
   1  &  $(1,1)$ &  292.1  &  1073  &  1237   &  148.3 & 1.177 \\ \hline
   2  &  $(2,1)$ &  465.3  &  1274  &  432.8  &  0     & 1.740 \\ \hline
   2  &  $(1,2)$ &  572.6  &  3049  &  2550   &  941.2 & 1.310 \\ \hline
   3  &  $(3,1)$ &  647.7  &  2407  &  416.6  &  0     & 2.293 \\ \hline
   3  &  $(1,3)$ &  917.2  &  7357  &  5246   &  900.5 & 1.416 \\ \hline
   4  &  $(2,2)$ &  845.3  &  3260  &  977.5  &  0     & 1.756 \\ \hline
   4  &  $(4,1)$ &  860.3  &  5317  &  431.1  &  0     & 3.128 \\ \hline
   4  &  $(1,4)$ &  1312   &  20454 & 17974   &  2007  & 1.498 \\ \hline
 \end{tabular}
\caption{The (dimensionless) mass $M_{*}$ and the size $\tilde{a}$ of the knot soliton and non-vanishing and finite components $V_{11}^{*}$, $U_{33}^{*}$, $W_{12}^{*}$ of the inertial tensor for various Hopf indices $Q_H=mn$ at $\Lambda_{\rm F}=1$ and $C_4=1$ for the Faddeev model.
}
\label{table:numerical-results}
\end{center}
\end{table}

Our numerical calculations  show
$M_{*} = 465$,  
$V_{11}^{*} = 1274$, 
$U_{33}^{*} = 433$ 
for 
 ($m=1, n=2; Q_H =2$).
By combining these results with those of numerical simulations of Yang-Mills theory on a lattice \cite{Teper98}:
$ 
  M_{2^{++}}/M_{0^{++}}  = 1.46 \pm 0.09 , 
$ 
a parameter is determined to be 
$C_4=[3/(M_{*}V_{11}^{*}(M_{2^{++}}/M_{0^{++}}-1))]^{1/2}$:
\begin{align}
 C_4=0.00303 \sim 0.00369  .
\end{align}
The ratio of the lowest glueball mass to the gauge-invariant gluon mass \cite{Kondo04}  due to (\ref{Faddeev-model-parameter}):
\begin{align}
 M_X \cong \sqrt{g^2 \langle -\mathbb{X}_\mu^2 \rangle_{\rm YM}} \cong \Lambda_{\rm F} C_4^{-1/2}  ,
\end{align}
is determined  from the ratio $M_{2^{++}}/M_{0^{++}}$ alone: 
\begin{align}
  M_{0^{++}}/M_X \cong C_4 M_{*} = 1.41 \sim 1.72 .
\end{align}
This results supports that a glueball is composed of  gluons. 
Another parameter $\Lambda_{\rm F}$ is determined from an input mass $M_{0^{++}}=(3.74\pm 0.12) \sqrt{\sigma}  \sim 1.5 {\rm GeV}$ to be
\begin{align}
 \Lambda_{\rm F} =  M_{0^{++}}/(C_4^{1/2}M_{*})= 0.0530 \sim 0.0586 {\rm GeV} .
\end{align}
The radius of the torus is 
$a= \Lambda_{\rm F}^{-1} C_4^{1/2} \tilde{a}=1.64 \sim 1.99~
{\rm (GeV)}^{-1}= 0.32 \sim 0.39~{\rm fm}$, 
which is the associated Compton wavelength to be compared to that of the pion with mass $m_\pi=140~{\rm MeV}$, $0.197/0.14=1.41~{\rm fm}$. 
This result is consistent with the prediction of phenomenological approaches, see e.g., \cite{potential-model}. 
Thus we obtain  an estimate of  the gluon mass
\begin{align}
 M_X  = 0.87 \sim 1.41~{\rm GeV} ,
\end{align}
and  the numerical value of the dimension two condensate:
\begin{align}
 \langle -\mathbb{X}_\mu^2 \rangle_{\rm YM} \cong \Lambda_{\rm F}^2 = (0.0530 \sim 0.058 {\rm GeV})^2  .
\end{align}

All the cases considered are summarized in Table~\ref{table:final-results}. 
The resulting values for $M_X$ are consistent with the result of numerical simulations on a lattice \cite{SIKKMS06}:
\begin{align}
 M_X  = 1.2~{\rm GeV} .
\end{align}

\begin{table} 
\small
\begin{center}
 \begin{tabular}{|c|c|c|c|c|c|c|}
  \hline
  \multicolumn{7}{|c|}{SU(2) Faddeev model}  \\ \hline
   $Q_H$  &  $(n,m)$ & $C_4$   & $C_4 M_{*}=M_{0^{++}}/M_X$  & $\Lambda_{\rm F}{\rm [GeV]}$ & $a {\rm [fm]}$ & $M_X {\rm [GeV]}$ \\ \hline
   1  &  $(1,1)$ &  0.00456  &  1.332  &  0.0760   &  0.206  & 1.126 \\ \hline
   2  &  $(2,1)$ &  0.00332  &  1.543  &  0.0560  &  0.352     & 0.972 \\ \hline
   2  &  $(1,2)$ &  0.00193  &  1.107  &  0.0596   &  0.190 & 1.356 \\ \hline
 \end{tabular}
\caption{The parameters of the Faddeev model determined from the identification with the glueball masses.  Only the central values are indicated.
}
\label{table:final-results}
\end{center}
\end{table}

\section{Conclusion and discussion}

We have performed the collective coordinate quantization of the classical knot soliton solution obtained under the toroidal ansatz of the Faddeev model. 
Then we have identified the resulting energy spectrum with the glueball mass spectrum of the SU(2) Yang-Mills theory, supposing the Faddeev model is a low-energy effective theory of SU(2) Yang-Mills theory.  
We have used two input masses $M_{0^{++}}$ and $M_{2^{++}}$ obtained by  numerical simulations of Yang-Mills theory on a lattice to determine two parameters in the Faddeev model for predicting  glueball masses other than $0^{++}$ and $2^{++}$. 

Thanks  to the NLCV, we can expect that the gauge-invariant mass term 
$\frac12 M_X^2 \mathbb{X}_\mu^2$ can be induced in Yang-Mills theory, since the dimension two composite operator 
$\mathbb{X}_\mu^2$ can be defined in the gauge-invariant way based on the NLCV. From the collective coordinate quantization of the Faddeev model and the identification of the glueball mass spectrum, we have obtained the reasonable value for the gauge-invariant gluon mass $M_X \propto \sqrt{g^2\left< -\mathbb{X}_\mu^2 \right>_{\rm YM}}$. 
These results are consistent with an observation that the gauge-invariant vacuum condensation $\left< \mathbb{X}_\mu^2 \right>_{\rm YM}$  of mass-dimension two takes place in Yang-Mills theory and gives a gluon mass $M_X \propto \sqrt{g^2\left< - \mathbb{X}_\mu^2 \right>_{\rm YM}}$.
The resulting mass scale enables us to derive the kinetic term in the Faddeev model as a LEET of Yang-Mills theory, which is needed to avoid the no-go theorem of Derrick and to allow the existence of the topological soliton as a result of   balancing the kinetic term  with the Skyrme term of four derivatives. 
Thus the dimension two condensate $\left< \mathbb{X}_\mu^2 \right>_{\rm YM}$ in QGD leads to the Faddeev model
and  the Faddeev model could be a realistic low-energy effective theory of SU(2) Yang-Mills theory.

It has been also shown \cite{KKMSSI06} that magnetic monopoles in QGD can be defined in the gauge-invariant way based on the NLCV.  
Then the Wilson loop average can be calculated from the linking number between the magnetic glue-knot and the electric Wilson loop.  
This will shed more light on the existence of mass gap in QGD and quark confinement in QCD. 
Extending  this work to color SU(3) case will be done in a subsequent paper.

\section*{Acknowledgments}

The authors would like to express their sincere thanks to Nobuyuki Sawado for private communications and for bringing their attention to \cite{KS05}, and Minoru Hirayama for illuminating conversations. 
This work is financially supported by 
Grant-in-Aid for Scientific Research (C) 18540251 from 
JSPS
and in part by Grant-in-Aid for Scientific Research on Priority Areas (B)13135203 from 
MEXT.




\end{document}